\documentclass[10pt]{article}
\usepackage{amsmath}
\usepackage{amssymb}
\usepackage{graphics}
\usepackage{rotating}
\usepackage{cite}
\usepackage{color}
\usepackage{colortbl}
\usepackage{fancybox}
\usepackage{pstricks}
\usepackage{color, colortbl}

\definecolor{ay}{rgb}{0.91, 0.84, 0.42}
\definecolor{arsenic}{rgb}{0.23, 0.27, 0.29}
\definecolor{darkseagreen}{rgb}{0.21, 0.37, 0.23}
\definecolor{darkred}{rgb}{0.55, 0.0, 0.0}
\definecolor{asparagus}{rgb}{0.53, 0.66, 0.42}
\definecolor{med}{rgb}{0.79, 0.86, 0.54}
\definecolor{arylideyellow}{rgb}{0.91, 0.84, 0.42}
\def\0{\mbox{\tiny $0$}}
\def\1{\mbox{\tiny $1$}}
\def\2{\mbox{\tiny $2$}}
\def\3{\mbox{\tiny $3$}}
\def\4{\mbox{\tiny $4$}}
\def\5{\mbox{\tiny $5$}}
\def\6{\mbox{\tiny $6$}}
\def\7{\mbox{\tiny $7$}}
\def\8{\mbox{\tiny $8$}}
\def\9{\mbox{\tiny $9$}}
\title{\hspace*{0cm}\shadowbox{\fcolorbox{black}{ay} { {\color{arsenic}{ \large \bf \begin{tabular}{c}
EXPERIMENTAL CONFIRMATION OF THE TRANSVERSAL\\ SYMMETRY BREAKING IN LASER PROFILES
\end{tabular}}}}}}

\author{
\small  Silv\^ania A. Carvalho$^{\1}$,\,\, Stefano De Leo$^{\2}$\thanks{deleo@ime.unicamp.br},\,\, Jos\'e A. Oliveira-Huguenin$^{\3}$\,\,and\,\,Lad\'ario da Silva$^{\3}$ \\
\small $^{\1}$ Department of Mathematics, Physics and Computation, State University of Rio de Janeiro, Resende (Brazil)\\
\small $^{\2}$ Department of Applied Mathematics, State University of Campinas, Campinas (Brazil)\\
\small $^{\3}$ Department of Physics, Fluminense Federal University, Volta Redonda (Brazil) }
\date{\small
\fcolorbox{black}{med} {\color{darkred} $\bullet$ {\color{arsenic}{
{\small \bf Journal of Modern Optics 64, 280-287 (2017) {\color{darkred}{ $\bullet$}} } }}}}

\begin{document}


\maketitle

\vspace*{-.7cm}

\begin{abstract}
\noindent
The Snell phase effects on the propagation of optical beams through dielectric blocks have been matter of recent theoretical studies.  The effects  of this phase  on the laser  profiles have been tested in our experiment. The data show an excellent agreement with the theoretical predictions confirming the  axial spreading modification and the  transversal symmetry  breaking. The possibility to set, by rotating the dielectric blocks, different configurations allows to recover the transversal symmetry. Based on this experimental evidence, dielectric  blocks  can be used as alternative optical tools to control the beam profile.
\end{abstract}







\section*{\color{darkred} \normalsize I. INTRODUCTION}
Studies of the  optical beam phase  confirmed \cite{AJPBlock} or  suggested  deviations \cite{GHS1947,1948AP437,GHS2013,2013JMO,2014JO16,2014PRA90,2016PRA93} from the law of geometrical optics \cite{Wolf1999,Saleh2007}. For example, the first order Taylor expansion of the Snell phase represents an alternative method  to calculate, by the stationary phase method \cite{AJPBlock}, the optical path of light beams.  In a recent work \cite{2016JMOTSB}, the effects of the second order Taylor expansion of the Snell phase on the beam profile was
investigated and an analytical formula found for the transversal symmetry breaking and the axial spreading modification.  If such theoretical predictions were confirmed by an experimental analysis,  the use of  the Snell phase  in describing the optical beam propagation through dielectric blocks  is not a matter of elegance  in determining the optical path, but a necessity in correctly predicting the beam profile behavior. In this case, the importance of the Snell phase could not only be seen   in confirming the law of geometrical optics (first order term expansion) but also in predicting deviations from the original shape of our beam (second order term expansion).

The beam profile control finds important applications in many research fields. In the mode matching of optical cavities, the adjustment of the beam width  is crucial for a good luminous energy storage \cite{Saleh2007}. An optical parametric oscillator  needs a good mode matching in order to amplify the conversion of signal and  idler beams, and to minimize the loose of squeezing by noise introduction \cite{OPO1, OPO2}. For the second harmonic generation in photonics crystal wave-guides, the mode matching should be carefully observed \cite{OPO3}. Additionally, the trapping of charged and neutral atoms has played a fundamental role in atomic and molecular physics \cite{1999ODT}. The common trap used for storage devices at ultra-low energies is the dipole trap which relies on the electric dipole interaction with far-detuned lasers. The result is a structure to trap atoms in the region of high intensity of a focused beam \cite{HGPRA2005,HG2005B}. In this context, a controlled rotation of dielectric blocks sheds new light on fine tuning  the axial spreading factor and, consequently, tuning of these optical systems in general. Thus, a device capable of controlling the beam width in a simple and precise way is of great importance for laser applications.

The aim of this paper is to experimentally detect  the axial spreading modification and the breaking of transversal symmetry of  gaussian beams propagating through dielectrics blocks. Our paper is organized as follows. Section II gives a brief exposition of the theoretical predictions and contains the analytical formula used to describe the outgoing beam. It is not our intention to present all the aspects of the theory leading to the analytical formula. For details, we refer the reader to \cite{2016JMOTSB}.  Section III discusses the experimental data, which  show an excellent agreement with the theoretical predictions both for the transversal symmetry breaking and for the axial spreading modification.  This suggests the possibility  to control the laser beam profile by an appropriate dielectric blocks setup. The final section contains our conclusions and  suggestions for future investigations.

\section*{\color{darkred} \normalsize II. AXIAL SPREADING AND TRANSVERSAL SYMMETRY BREAKING}
Let us consider, an incoming gaussian beam whose intensity is given by
\begin{equation}
\label{inc}
I_{_{\rm inc}}(\boldsymbol{r})= I_{_0}\, \left|\,\frac{\mbox{w}_{\0}^{^2}}{4 \pi}\, \int^{^{+\infty}}_{_{-\infty}}\hspace*{-.5cm}\mbox{d}k_x\, \int^{^{+\infty}}_{_{-\infty}}\hspace*{-.5cm} \mbox{d}k_y\,\,g(k_x,\,k_y)\, \exp \left[\,i\,\left(k_x\,x+k_y\,y+k_z\,z\right)\right]\,\right|^{^{2}}\,\,,
\end{equation}
where $g(k_x,\,k_y)= \exp \left[-\left(k_{x}^{^2}\,+\,k_{y}^{^2}\right)\,\mbox{w}_{\0}^{^2} \,/\, 4\right]$ and
$k_z=\sqrt{k^{^2}-k_x^{^2}-k_y^{^2}}$ ($k=2\,\pi/\lambda$).

For $\mbox{w}_{\0} \gtrsim\lambda$, the beam divergence is relatively small and we can use the paraxial approximation, $k_z \approx k-(k_x^{^2}+k_y^{^2})/2k$, which allows to analytically integrate the expression for the intensity of the incident beam, Eq.\,(\ref{inc}), obtaining the well-known expression describing the propagation of free  gaussian beams
\begin{equation}
\label{eq:Iinc}
\mbox{I}_{_{\rm inc}}(\boldsymbol{r})  =  \displaystyle{\frac{\mbox{I}_{\0}\,\mbox{w}_{\0}^{^2}}{\mbox{w}^{^{2}}(z)}} \, \, \exp\left[\,-\, 2 \,\,\displaystyle{\frac{x^{^{2}}+ y^{^{2}}}{\mbox{w}^{^{2}}(z)}}\, \right]\,\,,
\hspace*{1cm}
{\rm w}(z)={\rm w}_{\0}\,\sqrt{1 + \left(\,\frac{\lambda\,z}{\pi {\rm w}_{0}^{^{2}}}\right)^{^{2}}}\,\,.
\end{equation}
After passing through $N$ dielectric blocks like those drawn in Fig.\,1(a) and $M$ dielectric blocks, clockwise rotated with respect to  the previous ones by an angle $\pi/2$ along the $z$ axis, see Fig.\,1(b),
\begin{equation}
g(k_x,k_y)\,\,\,\rightarrow\,\,\, T^{^{^{[\rm TE,TM]}}}\hspace*{-.85cm}(k_x,k_y)\,g(k_x,k_y) = \left|\,T^{^{^{[\rm TE,TM]}}}\hspace*{-.85cm}(k_x,k_y)\,\right|\,g(k_x,k_y)\,
\exp\{\,i\,[\,\phi_{_{_{_{_{\rm Snell}}}}}\hspace*{-.5cm}(k_x,k_y) + \phi_{_{_{_{_{\rm GH}}}}}^{^{^{[\rm TE,TM]}}}  \hspace*{-.85cm}(k_x,k_y)   \,]\,\}\,\,,
\end{equation}
where $T^{^{[\rm TE,TM]}}$ are the transmission coefficients obtained by solving the electromagnetic Maxwell wave equations in the presence of stratified media.  These coefficients contain the geometrical phase, $\phi_{_{\rm Snell}}$, obtained by imposing the continuity of the electromagnetic field at dielectric/air interfaces\cite{AJPBlock} and the Goos-H\"anchen phase,  $\phi_{_{\rm GH}}$, apprearing in the Fresnel coefficients for incidence greater than the critical one\cite{GHS1947}. Observing that the effects on the beam propagation caused by the GH phase with respect to the ones caused by the Snell phase are proportional to $\lambda/\overline{AB}$, for the purpose of our investigation,  without loss of generality, we can  only consider the Snell phase. The first order terms are clearly responsible for the transversal lateral displacements. Indeed, they directly act on the transversal spatial phase $\exp[\,i\,(\,k_x\, x +k_y\, y\,)\,]$
modifying the center of the incoming beam,
\begin{equation}
\{\,x_{_{\rm inc}}\,,\,y_{_{\rm inc}}\,\}=\{\,0\,,\,0\,\}
\,\,\,\,\,\,\,\rightarrow\,\,\,\,\,\,\, \{\,x_{_{\rm Snell}}\,,\,y_{_{\rm Snell}}\,\}   =  \left\{\,-\,\frac{\partial \phi_{_{\rm Snell}}}{\partial k_x}\,\,,\,  -\,\frac{\partial \phi_{_{Snell}}}{\partial k_y}   \,\right\}_{_{(0,0)}}\,\,,
\end{equation}
confirming the optical path predicted by the laws of geometrical optics\cite{AJPBlock}.  The Snell second order terms act on the axial spatial phase $\exp[\,-\,i\,(\,k_x^{^{2}} +k_y^{^{2}}\,)\,z/\,2\,k\,]$ modifying the axial spreading in the transversal plane $xz$ and $yz$\cite{2016JMOTSB},
 \begin{equation}
  \left\{\,\mbox{w}\left(z\right)\,\,,\,
 \mbox{w}\left(z\right)
 \,\right\}\,\,\,\,\,\,\,\rightarrow\,\,\,\,\,\,\, \left\{\,\mbox{w}\left(z -\,k\,  \frac{\partial^{^2} \phi_{_{\rm Snell}}}{\partial k_x^{^2}}\right)\,\,,\,
 \mbox{w}\left(z -\,k\,  \frac{\partial^{^2} \phi_{_{\rm Snell}}}{\partial k_y^{^2}}\right)
 \,\right\}_{_{(0,0)}}\,\,.
 \end{equation}
Consequently, propagation through dielectric blocks causes a symmetry breaking with respect to the free propagation and more important, as predicted in ref.\,\cite{2016JMOTSB}, a focalization like effect. The outgoing beam intensity, for incidence greater than the critical one, can then be expressed by
\begin{equation}
\label{eq:Iout}
\mbox{I}^{^{[\rm TE,TM]}} (\boldsymbol{r},\theta)  =  \displaystyle{\frac{\mbox{I}_{\0}\,\mbox{w}_{\0}^{\2}\,\left|T^{^{[\rm TE,TM]}}_{_{\theta}}\right|^{^{2\, (N+M)}}}{\mbox{w}(z - \, z^{^{\perp}}_{_{\rm Snell}})\,\mbox{w}(z - z_{_{\rm Snell}})}} \, \, \exp\left\{\,- 2 \left[\displaystyle{\frac{\left(x- x_{_{\rm Snell}}\right)^{^{2}}}{\mbox{w}^{^{2}}(z - z^{^{\perp}}_{_{\rm Snell}})}}\, + \displaystyle{\frac{\left(y - y_{_{\rm Snell}}\right)^{^{2}}}{\mbox{w}^{^{2}}(z - z_{_{\rm Snell}})}}\right]\,\right\} \,\,,
\end{equation}
where
\begin{equation}
\label{Fresnel}
\left\{\,\left|T^{^{[TE]}}_{_{\theta}}\right|\,,\, \left|T^{^{[TM]}}_{_{\theta}}\right|\,\right\} = \left\{\,
\frac{4\,n\,\cos\psi\,\cos\theta}{(\cos\theta\,+\,n\cos\psi)^{^2}}\,,\,
\frac{4\,n\,\cos\psi\,\cos\theta}{(n\,\cos\theta\,+\,\cos\psi)^{^2}}\,\right\}
\end{equation}
are the Fresnel transmission coefficients for transverse electric (TE) and transverse magnetic (TM) waves\cite{2013JMO}. The angle $\psi$ is obtained  from the incidence angle $\theta$ by the Snell law, $\sin\theta=n\,\sin \psi$, and the angle $\varphi$ (incidence at the down dielectric air interface) is given by $\psi+\frac{\pi}{4}$\cite{2013JMO}.
\begin{equation}
\left\{x_{_{\rm Snell}}\,,\,y_{_{\rm Snell}}\right\}  =  \left\{\,M\,\frac{\cos\theta - \sin\theta}{\sqrt{2}}\,,\,N\,\frac{\cos\theta - \sin\theta}{\sqrt{2}} \right\} \,\,\overline{BC}
\label{center}
\end{equation}
is the center of the beam obtained by the first order term of the Taylor expansion of the Snell phase (confirming the optical path predicted by the ray optics) \cite{AJPBlock}. $\overline{BC}=\sqrt{2}\,\tan\varphi\, \overline{AB}$ guarantees that the laser exit point from the dielectric block has the same height of the incoming one \cite{2014JO16,2014PRA90}, and, finally,
\begin{equation}
\left\{z^{^{\perp}}_{_{\rm Snell}}\,,\,z_{_{\rm Snell}}\right\}  =  \left\{\,M\,f_{_\theta} + N\,g_{_\theta} \,,\, M\,g_{_\theta} + N\,f_{_\theta} \right\}\,\overline{AB}\,\,,
\label{axial}
\end{equation}
with
\begin{eqnarray}
f_{_\theta} & =& (\sin\theta\,+\,\cos\theta)\,\tan\varphi\,-\,\frac{\cos^{^2}\theta}{n\,\cos^{^3}\psi}\,(1\,+\,
\tan\varphi)\,\,,\nonumber \\
g_{_\theta} & = &  \sin\theta\,+\,\cos\theta\,\tan\varphi\,-\,\frac{\cos^{^2}\theta}{n\,\cos\psi}\,
(1\,+\,\tan\varphi)\,\,,
\end{eqnarray}
gives the axial spreading modifications due to the second order contribution of the Snell phase expansion, for details see ref.  \cite{2016JMOTSB}.
For $N\neq M$,
\[z_{_{\rm Snell}}^{^{\perp}}\,\,\neq\,\,z_{_{\rm Snell}}\]
which causes  the transversal symmetry breaking.

Now that we are familiar with the mathematical formulation of the transmitted beam, it would be natural to consider a particular incidence angle to prepare the experiment. The choice of  $\theta = \pi/4$ (see Fig.\,1) has the advantage of
not modifying the center of the beam (\ref{center}),
\begin{equation}
\left\{x_{_{\rm Snell}}\,,\,y_{_{\rm Snell}}\right\} = \left\{0\,,\,0\right\}\,\,.
\end{equation}
For this incidence angle, we have
\begin{equation*}
\left\{\,\sin\psi\,,\,\cos\psi\,,\,\sin\varphi\,,\,\cos\varphi\,\right\}  =  \left\{\,\frac{1}{\sqrt{2}\,n}\,,\,\frac{\sqrt{2n^{^{2}} - 1}}{\sqrt{2}\,n}\,,\,\frac{\sqrt{2n^{^{2}} - 1} + 1}{2\,n}\,,\,\frac{\sqrt{2n^{^{2}} - 1} - 1}{2\,n} \right\}
\end{equation*}
and, consequently, the Fresnel coefficients (\ref{Fresnel}) become
\begin{equation}
\left\{\,\left|T^{^{[TE]}}_{_{\pi/4}}\right|\,,\, \left|T^{^{[TM]}}_{_{\pi/4}}\right|\,\right\} =
\left\{\,\frac{4 \sqrt{2 n^{\2}-1}}{\left(1+\sqrt{2 n^{\2}-1}\right)^{^2}}\,,\,\frac{4 n^{\2} \sqrt{2 n^{\2}-1}}{\left(n^{\2}+\sqrt{2 n^{\2}-1}\right)^{^2}}\,\right\}\,\,.
\end{equation}
It is possible to establish the axial modifications by knowing the number and the kind of blocks, ($M,N$), and  the factors
\begin{equation}
f_{_{\pi/4}} = \displaystyle{\sqrt{2}\,\left(2n^{^{2}} + \sqrt{2n^{^{2}} - 1}\right)\big{/}\left(2n^{^{2}} - 1\right)}\hspace{0.8cm}\mbox{and}\hspace{0.8cm} g_{_{\pi/4}} = \sqrt{2}\,\,.
\label{eq:eqfg}
\end{equation}
The normalized intensity,
\[
\mathcal{I}\,(\boldsymbol{r},\theta) =  \frac{ \mbox{I}^{^{[\rm TE,TM]}} (\boldsymbol{r},\theta) \mbox{w}(z - \, z^{^{\perp}}_{_{\rm Snell}})\,\mbox{w}(z - z_{_{\rm Snell}})}{\mbox{I}_{\0}\,\mbox{w}_{\0}^{^2}\,\left|T^{^{[\rm TE,TM]}}_{_{\theta}}\right|^{^{2\, (N+M)}}}\,\,,
\]
is clearly independent of the polarization and for the incidence angle $\pi/4$ becomes
\begin{equation}
\mathcal{I}\,(\boldsymbol{r},\mbox{$\frac{\pi}{4}$}) = \exp\left\{- \,2 \left[\,
\displaystyle{\frac{x^{^{2}}}{\mbox{w}^{^{2}}[z - (\,M\,f_{_{\pi/4}}+N\,g_{_{\pi/4}})\,\overline{AB}\,]}}\, +
\displaystyle{\frac{y^{^{2}}}{\mbox{w}^{^{2}}[z - (\,M\,g_{_{\pi/4}}+N\,f_{_{\pi/4}})\,\overline{AB}\,]}}
\,\right]\right\}\,\,.
\label{outB}
\end{equation}
The choice of a configuration with a number of blocks  $M=N$  allows to recover the transversal symmetry. In this case,  the outgoing beam will be a symmetric gaussian beam  and its width will be reduced with respect to the width of a beam propagating in the free space (focalization like effect caused by the propagation through the dielectric blocks).

\section*{\color{darkred} \normalsize III. EXPERIMENTAL EVIDENCE OF TRANSVERSAL SYMMETRY BREAKING}

Before discussing the experimental data, let us describe the experimental setup used
to observe the transversal symmetry breaking and the axial spreading modification. The  incoming beam  was  a TM polarized gaussian DPSS laser (the choice of a particular polarization does not influence the profile modifications)  with a  power of $1,5$\,mW, with a wavelength
\[ \lambda=532\,{\rm nm}\,\,, \]
 and  focused, by a lens of focal length $f=50$\,cm, in order to obtain a beam waist radius
 \[\mbox{w}_{\0} = (\,79.0 \pm 1.5\,) \,\mu {\rm m}\,\,.\]
 The beam parameters were determined by applying the knife-edge method along the beam path \cite{KEM2006}.
  Recalling that
 \begin{equation*}
 {\rm w}(z)={\rm w}_{\0}\,\sqrt{1 + \left(\,\frac{\lambda\,z}{\pi {\rm w}_{0}^{^{2}}}\right)^{^{2}}}
 \end{equation*}
and observing that
\begin{eqnarray}
\label{errw}
\frac{\sigma[ {\rm w}(z)]}{{\rm w}(z)} & = & \sqrt{\left[\,\frac{\partial {\rm w}(z)}{\partial {\rm w}_{\0}}\,\,\sigma({\rm w}_{\0})\,\right]^{^{2}}+ \left[\,\frac{\partial {\rm w}(z)}{\partial z}\,\,\sigma(z)\,\right]^{^{2}}}\,\,\mbox{\huge $/$}\,{\rm w}(z)
\nonumber\\
 & = &
\sqrt{\left[\,2\,\frac{{\rm w}_{\0}}{{\rm w}(z)} - \, \frac{{\rm w}(z)}{{\rm w}_{\0}}\,\right]^{^{2}}\,\sigma^{^{2}}({\rm w}_{\0}) + \left[\, 1 - \,\frac{{\rm w}_{\0}^{^{2}}}{{\rm w}^{^{2}}(z)} \,\right]\,\left(\,\frac{\lambda}{\pi\,{\rm w}_{\0}}\,\right)^{^{2}}\,\sigma^{^{2}}(z)}\,\,\mbox{\huge $/$}\,{\rm w}(z)    \nonumber \\
 & \approx & \left[\,1\,-\,2\,\frac{{\rm w}^{^{2}}_{\0}}{{\rm w}^{^{2}}(z)} \,\right]\,\frac{\sigma({\rm w}_{\0})}{{\rm w}_{\0}}\hspace*{2cm}
 [\,{\rm w}(z)>\sqrt{2}\,{\rm w}_{\0}\,]\,\,,
\end{eqnarray}
for free propagation, at the power meter located at
\[  z_{_{\rm PM}} = (\,39.5 \pm 0.1\,)\,{\rm cm}\,\,,\]
we should  find the following beam width
\begin{equation}
{\rm w}_{_{\rm Teo}}^{^{\rm Free}} = (\,850.4 \pm 15.9\,) \,\mu {\rm m}\,\,.
\end{equation}
Each dielectric block used in the experiment had a refractive index  $n=1.515$ (BK7) and dimensions
\[
\underbrace{(\,91.5\pm 0.5\,)\,{\rm mm}}_{\overline{BC}}\,\, \times\,\,
\underbrace{(\,20.0\pm 0.5\,)\,{\rm mm}}_{\overline{AB}}\,\, \times\,\,\,
(\, 14.0\pm 0.5\,)\,{\rm mm}
\]
see Fig.\,1(a-b). The incoming beam propagates along the $z$-axis (see Fig.\,1) and  forms  with the left interface an angle
of $\theta=\pi/4$ in both  the (a) and (b) configurations. To guarantee that the transmitted beam coming out of  the dielectric block at the same height of the incident point,
\[   \overline{BC} = \sqrt{2}\,\,\frac{\sqrt{2\,n^{^{2}}-1}+1}{\sqrt{2\,n^{^{2}}-1}-1}\,\,\,\overline{AB}\,\,.\]
The blocks faces are anti-reflection coated to reduce unwanted internal reflections.

In order to investigate the axial spreading modification and the transversal symmetry breaking, we used two different block configurations. Fig.\,1(a) shows a block in which the plane of propagation is the $y$-$z$ plane. This dielectric structure  corresponds to the configuration $(M,N) = (0,1)$.  By rotating the block  clockwise along the $z$-axis with  an angle $\pi/2$, we  get a dielectric structure in which the plane of propagation is  the $x-z$ plane, as shown in Fig.\,1(b). This dielectric structure corresponds to the   configuration $(M,N) = (1,0)$. In Fig. 1(c), we have the configuration $(M\,,\,N) = (2\,,\,1)$, i.e.  two blocks with a plane of propagation in the $x-z$ plane and one with a plane of propagation in the $y-z$ plane. The order is not relevant and the transversal profiles  will be the same for all possible permutations of the blocks. The symmetry is recovered after the beam passes through the first two blocks with a different plane of propagation. After passing through the last block the transversal symmetry breaking acts principally on the $x$-axis reverting  what happened after the first block. By using
\begin{equation}
\left(\,{\rm w}_{_{\rm Teo}}^{^{{\rm Snell},\perp}}\,,\,{\rm w}_{_{\rm Teo}}^{^{{\rm Snell}}}\,\right) =
\left(\,{\rm w}[z_{_{\rm PM}} - (\,M\,f_{_{\pi/4}}+N\,g_{_{\pi/4}})\,\overline{AB}\,] \,,\,{\rm w}[z_{_{\rm PM}} - (\,M\,g_{_{\pi/4}}+N\,f_{_{\pi/4}})\,\overline{AB}\,] \,\right)\,\,,
\end{equation}
we can give the theoretical predictions of the outgoing beam widths at the power meter
\begin{equation}
\label{TeoW}
\left(\,{\rm w}_{_{\rm Teo}}^{^{{\rm Snell},\perp}}\,,\,{\rm w}_{_{\rm Teo}}^{^{{\rm Snell}}}\,\right) =
\left\{
\begin{array}{cccr}
(\,669.5\pm 12.4\,,\,524.2\pm\,\,\, 9.5\,) & \mu{\rm m} & \hspace*{1cm} & (\,M\,,\,N\,) =  (\,0\,,\, 3\,)\\
(\,681.2\pm12.6\,,\,681.2\pm 12.6\,) & & \hspace*{1cm} & (\,1\,,\, 1\,)
\end{array}
\right.
\,\,,
\end{equation}
where the errors were obtained by using Eq.\,(\ref{errw}).

 The experimental widths were calculated by
 \begin{equation}
\left(\,{\rm w}_{_{\rm Exp}}^{^{{\rm Snell},\perp}}\,,\,{\rm w}_{_{\rm Exp}}^{^{{\rm Snell}}}\,\right) =  \left(\,
x_{_{\rm KM}}\,,\,y_{_{\rm KM}}\,\right)\,/\,\alpha\,\,,
\end{equation}
where  $x_{_{\rm KM}}$ and $y_{_{\rm KM}}$ were found
by the knife-method for
  \[
  2.275\%{\rm -}97.725\%\,\,\,\,\,{\rm (a)}\,\,,\,\,\,\,\,\,\,
   5\%{\rm -}95\%\,\,\,\,\,{\rm (b)}\,\,,\,\,\,\,\,\,\,
   10\%{\rm -}90\%\,\,\,\,\,{\rm (c)}\,\,,\,\,\,\,\,\,\,
    20\%{\rm -}80\%\,\,\,\,\,{\rm (d)}\,\,,
    \]
 and the $\alpha$ factor obtained by
 \[
 \sqrt{\frac{2}{\pi}}\,\int^{^{+\,\alpha}}_{_{-\,\alpha}}\,\hspace*{-.3cm}\mbox{d}r\,\,e^{-\,2\,r^{\2}}=
 95.45\%\,\,\,\,\,{\rm (a)}\,\,,\,\,\,\,\,\,\,
 90\%\,\,\,\,\,{\rm (a)}\,\,,\,\,\,\,\,\,\,
 80\%\,\,\,\,\,{\rm (a)}\,\,,\,\,\,\,\,\,\,
 60\%\,\,\,\,\,{\rm (a)}\,\,,\,\,\,\,\,\,\,
 \]
 which implies
  \[
\alpha=  1\,\,\,\,\,{\rm (a)}\,\,,\,\,\,\,\,\,\,
  0.82\,\,\,\,\,{\rm (b)}\,\,,\,\,\,\,\,\,\,
  0.64\,\,\,\,\,{\rm (c)}\,\,,\,\,\,\,\,\,\,
  0.42\,\,\,\,\,{\rm (d)}\,\,.
  \]
The experimental data for $x_{_{\rm KM}}$ and $y_{_{\rm KM}}$ and the corresponding beam widths  ${\rm w}_{_{\rm Exp}}^{^{{\rm Snell},\perp}}$ and ${\rm w}_{_{\rm Exp}}^{^{{\rm Snell}}}$ are listed in Table 1.1. The experimental widths are in agreement with the theoretical predictions given in Eq.\,(\ref{TeoW}).

By using the measurements done by the knife-method, we can also obtain the intensity at the power meter,
\begin{equation}
\label{ICamExp}
I_{_{\rm Exp}}(x,y) =\exp\left\{\,-\, 2\,\left[\,
\left(\, \frac{x_{_{\rm KM}}}{{\rm w}_{_{\rm Teo}}^{^{{\rm Snell},\perp}}}\,\right)^{^{2}}\,+\,
\left(\, \frac{y_{_{\rm KM}}}{{\rm w}_{_{\rm Teo}}^{^{{\rm Snell}}}}\,\right)^{^{2}}
 \, \right]\,\right\}\,\,,
\end{equation}
 and its errors,
\begin{eqnarray}
\frac{\sigma[I_{_{\rm CAM}}^{^{\rm Exp}}(x,0)]}{I_{_{\rm CAM}}^{^{\rm Exp}}(x,0)} &=& 4\,\frac{x}{{\rm w}_{_{\rm CAM}}^{^{{\rm Snell},\perp}}}\,\,\sqrt{
\left[\,\frac{\sigma(x)}{{\rm w}_{_{\rm CAM}}^{^{{\rm Snell},\perp}}}\,\right]^{^{2}}\,+\,
\left[\,\frac{x\,\sigma({\rm w}_{_{\rm CAM}}^{^{{\rm Snell},\perp}})}{{\rm w}_{_{\rm CAM}}^{^{{\rm Snell},\perp\,\,\,2}}}\,\right]^{^{2}}}\,\,,\nonumber \\
\frac{\sigma[I_{_{\rm CAM}}^{^{\rm Exp}}(0,y)]}{I_{_{\rm CAM}}^{^{\rm Exp}}(0,y)} &=& 4\,\frac{y}{{\rm w}_{_{\rm CAM}}^{^{{\rm Snell}}}}\,\,\sqrt{
\left[\,\frac{\sigma(y)}{{\rm w}_{_{\rm CAM}}^{^{{\rm Snell}}}}\,\right]^{^{2}}\,+\,
\left[\,\frac{y\,\sigma({\rm w}_{_{\rm CAM}}^{^{{\rm Snell}}})}{{\rm w}_{_{\rm CAM}}^{^{{\rm Snell}\,\,\,2}}}\,\right]^{^{2}}}\,\,.
\end{eqnarray}
These intensity values are listed in Table 1.2 and were used in the plots of Fig.\,2, configuration $(M,N)=(0,3)$, and Fig.\, 3, configuration $(1,1)$, to compare the experimental data with
the theoretical curves (blue solid lines) obtained by Eq.\,(\ref{outB}). The agreement is excellent and confirms the theoretical predictions. The dashed black lines represent the beam intensity  for free propagation. As can be seen in all three configurations, we have a clear spreading delay of the laser passing through dielectric blocks. In Fig.\,2, we can see a maximal  breaking of symmetry between the transversal coordinates. The symmetry is recovered in Fig.\,3 where $M=N=1$.

\section*{\color{darkred} \normalsize IV. CONCLUSIONS}

The Goos-H\"anchen phase has been the matter of studies since the fifties \cite{GHS1947,1948AP437}. The Snell phase, often forgotten, has recently been  used as an alternative way to obtain the optical path predicted by geometrical optics \cite{AJPBlock} and, even  more important, to predict the axial spreading modification and the breaking of symmetry in the transversal profiles of laser propagating through dielectric blocks \cite{2016JMOTSB}.

In this paper, we have experimentally tested the theoretical predictions, for an incidence angle of $\pi/4$,  and found an excellent agreement with the analytical formula given by literature \cite{2016JMOTSB}. This excellent agreement of  experimental data and theory can be regarded as a convincing proof of the importance to  use  the Snell phase in order to correctly  describe the propagation of optical beams through dielectric blocks, on one hand, and of the possibility to use dielectric structures to control the laser beam profile, on the other hand. Indeed,  this profile modification technique, allowing to focus the beam and to act on its transversal symmetry,  is better than using a simple lens.

Future studies  could test the formula at different angles and observe, in particular, the results   in the proximity of the critical angle where analytical formulas cannot be obtained and where the Goos-H\"anchen phase could play a relevant role. In this case, the experimental data should be compared with numerical calculations. We also expect phenomena of transversal symmetry breaking in Herimite and Laguerre-Gauss beams propagating through dielectric blocks.

\vspace*{0.5cm}

\noindent \textbf{\footnotesize \color{darkred} ACKNOWLEDGEMENTS}\\
The authors thank the CNPq, FAPERJ and PROPPI-UFF for the financial support. They also  thank Prof.\,Luis Ara\'ujo e Dr.\,Manoel P. Ara\'ujo for useful comments and stimulating conversations and Dr.\,Rita K. Kraus for a critical reading of the manuscript. The authors are greatly indebted with an anonymous referee who, with his suggestions and observations, helped to improve the final version of the paper.

\newpage

\begin{table}
\vspace*{-5cm}
\centering
\caption{In Table 1.1, we find the experimental beam widths  obtained by the knife-method measurements done for power cuts  of $2.275\%$-$97.725\%$ (a),  $5.0\%$-$95.0\%$ (b), $10.0\%$-$90.0\%$ (c), and $20.0\%$-$80.0\%$ (d). The data show agreement with the theoretical widths. In Table 1.2, we give the experimental values of the transversal intensities, calculated by using Eq.\,(\ref{ICamExp}). The error in the $x_{_{\rm KM}}$ and $y_{_{\rm KM}}$ measurement is of $15\mu {\rm m}$.}
\vspace*{1cm}
\shadowbox{{\bf Table 1.1}}\\
\vspace*{0.5cm}
\begin{tabular}{|c||c||c|c|c||c|c|c|}
\hline
\rowcolor{med}
KM  &
$(M\,,\,N)$ &
$x_{_{\rm KM}}[\,\mu{\rm m}]$ &
${\rm w}_{_{\rm Exp}}^{^{{\rm Snell},\perp}}[\,\mu{\rm m}]$  &
 ${\rm w}_{_{\rm Teo}}^{^{{\rm Snell},\perp}}[\,\mu{\rm m}]$  &
$y_{_{\rm KM}}[\,\mu{\rm m}]$ &
${\rm w}_{_{\rm Exp}}^{^{{\rm Snell}}}[\,\mu{\rm m}]$ &
${\rm w}_{_{\rm Teo}}^{^{{\rm Snell}}}[\,\mu{\rm m}]$  \\
\hline
\hline
\rowcolor{ay!10}
  (a)   &   $(0,3)$  &   $685$ &  $685.0 \pm 15.0$ & $669.5 \pm  12.4$ &  $545$ &  $545.0 \pm 15.0$  & $524.2 \pm \,\,9.5$ \\
\rowcolor{ay!10}
     &   $(1,1)$  &   $695$ &  $695.0 \pm 15.0$ & $681.2 \pm 12.6$ &  $690$ &  $690.0 \pm 15.0$  & $681.2 \pm 12.6$ \\
\hline
\hline
\rowcolor{ay!40}
 (b)    &   $(0,3)$  &   $550$ &  $670.7 \pm 18.3$ & $669.5 \pm  12.4$ &  $440$ &  $536.6 \pm 18.3$  & $524.2 \pm \,\,9.5$ \\
\rowcolor{ay!40}
     &   $(1,1)$  &   $555$ &  $676.8 \pm 18.3$ & $681.2 \pm 12.6$ &  $550$ &  $670.7 \pm 18.3$  & $681.2 \pm 12.6$\\
\hline
\hline
 \rowcolor{ay!70}
 (c)    &   $(0,3)$  &   $425$ &  $664.1 \pm 23.4$ & $669.5 \pm  12.4$ &  $340$ & $531.3 \pm 23.4$  & $524.2 \pm \,\,9.5$\\
\rowcolor{ay!70}
     &   $(1,1)$  &   $430$ &  $671.9 \pm 23.4$ & $681.2 \pm 12.6$ &  $430$ &  $671.9 \pm 23.4$  & $681.2 \pm 12.6$\\
\hline
\hline
\rowcolor{ay!100}
 (d)    &   $(0,3)$  &   $275$ &  $654.8 \pm 35.7$ & $669.5 \pm  12.4$ &  $225$ &  $535.7 \pm 35.7$  & $524.2 \pm \,\,9.5$\\
\rowcolor{ay!100}
     &   $(1,1)$  &   $275$ &  $654.8 \pm 35.7$ & $681.2 \pm 12.6$ &  $285$ &  $678.6 \pm 35.7$  & $681.2 \pm 12.6$\\
\hline
\end{tabular}
\\
\vspace*{1cm}
\shadowbox{{\bf Table 1.2}}\\
\vspace*{0.5cm}
\begin{tabular}{|c||c||c|c||c|c|}
\hline
\rowcolor{med}
KM  & $(M\,,\,N)$ & $x_{_{\rm KM}}[\,\mu{\rm m}]$ & $I_{_{\rm Exp}}(x,0)\,[\%]$ & $y_{_{\rm KM}}[\,\mu{\rm m}]$ &
$I_{_{\rm Exp}}(0,y)\,[\%]$\\
\hline
\hline
\rowcolor{ay!10}
  (a)    &   $(0,3)$ &   $685$ &    $12.3 \pm 1.5$ &  $545$ &   $11.5 \pm 1.6$ \\
\rowcolor{ay!10}
      &   $(1,1)$ &   $695$ &   $12.5 \pm 1.5$ &   $690$ &   $12.8 \pm 1.5$ \\
\hline
\rowcolor{ay!40}
 (b)     &   $(0,3)$ &   $550$ &   $25.9 \pm 2.3$ &   $440$ &   $24.4 \pm 2.7$ \\
\rowcolor{ay!40}
    	&   $(1,1)$ &   $555$ &   $26.5 \pm 2.3$ &   $550$ &   $27.1 \pm 2.3$\\
\hline
\rowcolor{ay!70}
 (c)     &   $(0,3)$ &   $425$ &   $44.7 \pm 2.9$ &   $340$ &   $43.1 \pm 3.5$\\
 \rowcolor{ay!70}
    	&   $(1,1)$ &   $430$ &   $45.1 \pm 2.8$ &   $430$ &   $45.1 \pm 2.8$ \\
\hline
\rowcolor{ay!100}
 (d)     &   $(0,3)$  &   $275$ &   $71.4 \pm 2.8$ &   $225$ &   $69.2 \pm 3.5$ \\
\rowcolor{ay!100}
      &   $(1,1)$ &   $275$ &   $72.2 \pm 2.7$ &   $285$ &   $70.5\pm 2.8$ \\
\hline
\end{tabular}
\end{table}

\newpage

\begin{figure}
\vspace*{-1.5cm} \hspace*{-1.cm}
\includegraphics[width=17.5cm, height=21cm, angle=0]{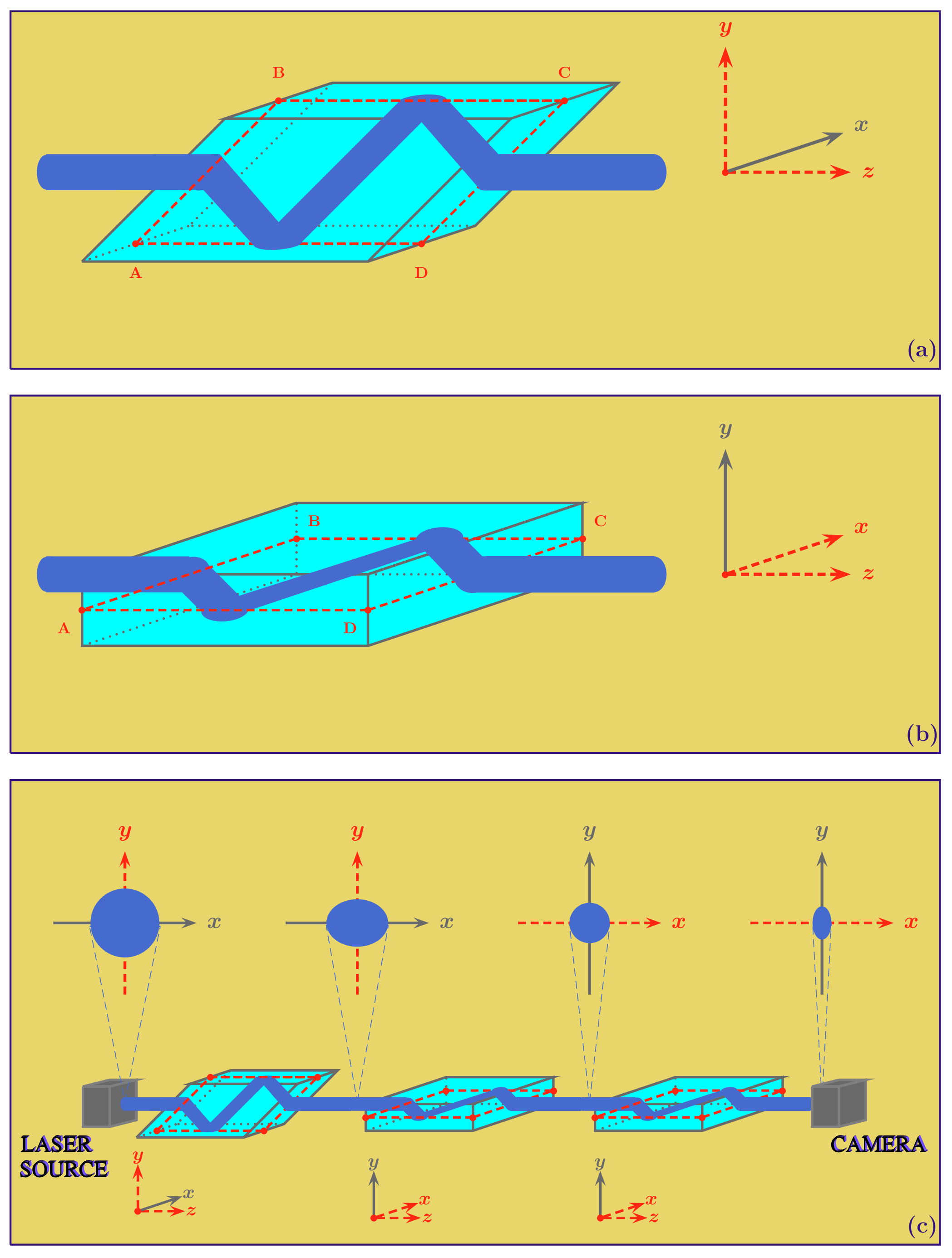}
\vspace*{0.cm}
 \caption{ Experimental setup. The incoming beam propagate along the $z$-axial forming an angle of $\pi/4$ with the left interface of the dielectric blocks.  In (a), the plane of propagation is the plane $y$-$z$ and represents the  configuration $M,N)=(0,1)$. In (b), the block is clockwise rotated by an angle  $\pi/2$ along the axial axis, configuration $(M,N)=(1,0)$. In $(c)$, we have the configuration $(M,N)=(2,1)$. The transversal symmetry is recovered when the beam passes through the same number of blocks $M$ and $N$. }
\label{fig:Fig1}
\end{figure}

\newpage

\begin{figure}
\vspace*{-2cm} \hspace*{-2.5cm}
\includegraphics[width=19cm, height=22cm, angle=0]{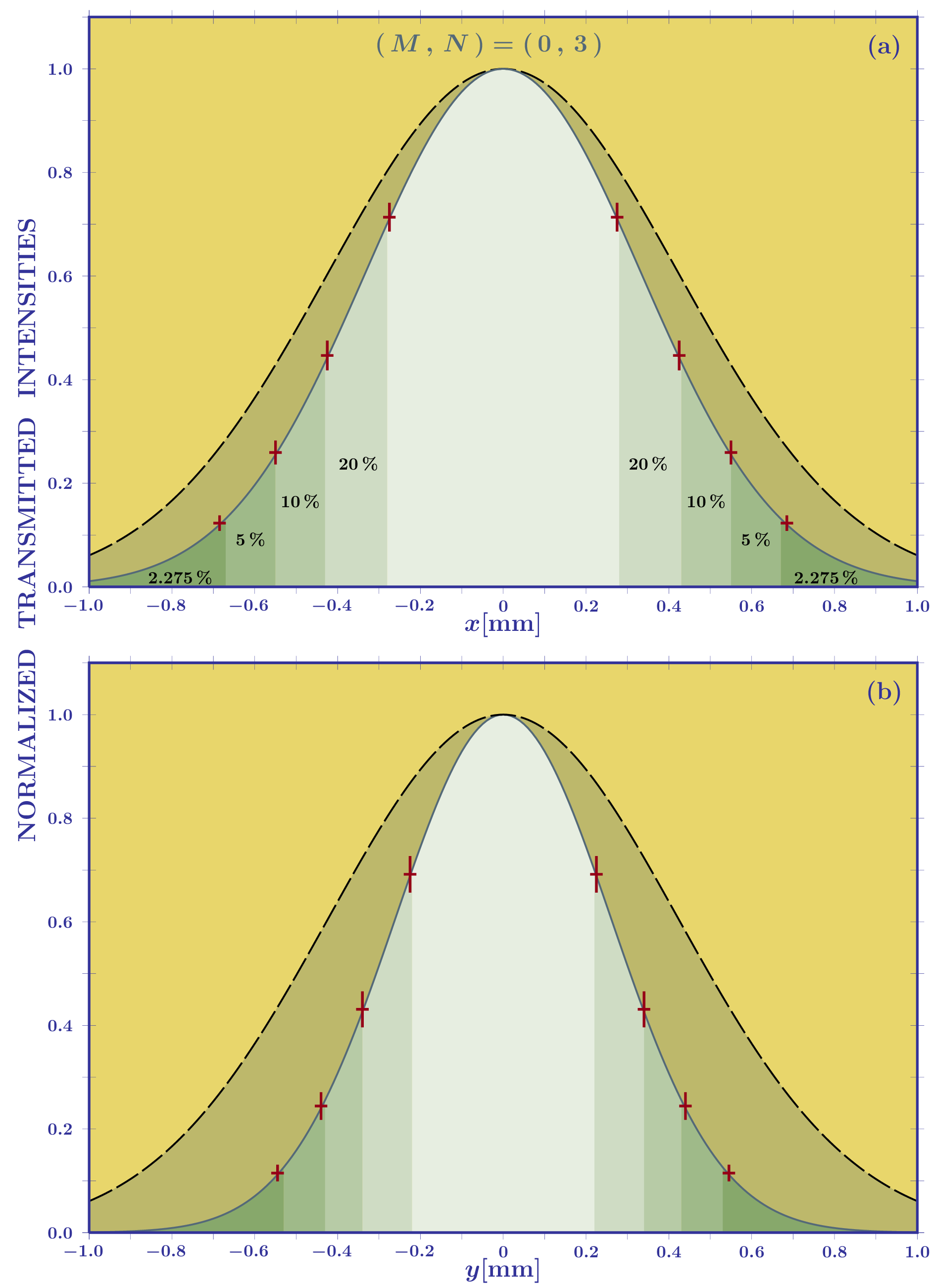}
\vspace*{-0cm}
 \caption{Transversal profiles of the normalized transmitted intensity. The dashed black lines represent the profiles for free propagation. The blue solid lines are the profiles for beams passing through the dielectric configuration   $(M,N)=(0,3)$. The breaking of symmetry between the $x$ and $y$ axis  is clear and the axial spreading delay  with respect to the free propagation is more evident for the transversal component $y$. The experimental data show an excellent agreement with the theoretical curves.}
\label{fig:Fig2}
\end{figure}

\newpage

\begin{figure}
\vspace*{-2cm} \hspace*{-2.5cm}
\includegraphics[width=19cm, height=22cm, angle=0]{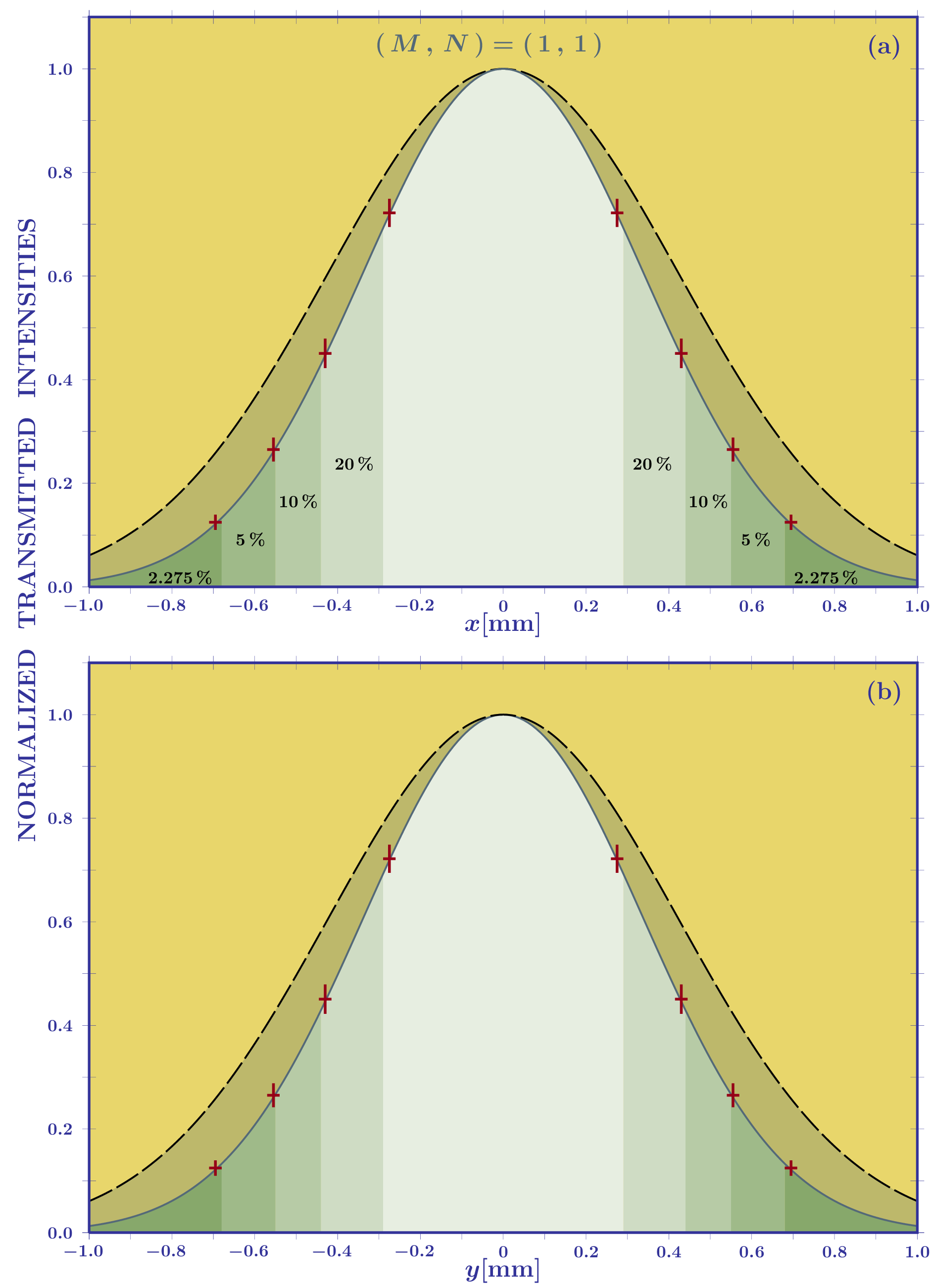}
\vspace*{-0cm}
 \caption{Transversal profiles of the normalized transmitted intensity. The dashed black lines represent the profiles for free propagation. The blue solid lines are the profiles for beams passing through the dielectric configuration   $(M,N)=(1,1)$. The symmetry between the transversal components $x$ and $y$ is recovered. It is also possible to see the axial spreading delay  with respect to the free propagation. The experimental data show an excellent agreement with the theoretical curves.}
\label{fig:Fig3}
\end{figure}


\begin{thebibliography}{100}



\bibitem{AJPBlock}
S.\,A. Carvalho and S. De Leo,
{\em The use of the stationary phase method as a mathematical tool to determine the path of optical beams},
Am.\,J.\,Phys. \textbf{83}, 249-255 (2015).




\bibitem{GHS1947}
F. Goos and H. H\"anchen,
{\em Ein neuer und fun\-da\-men\-ta\,ler Versuch zur total\-re\-fle\-xi\-on},
Ann.\,der\,Physik {\bf 436}, 333-346 (1947).

\bibitem{1948AP437}
K. Artmann,
{\em Berechnung der Seitenversetzung des Totalreflektierten Strahles},
Annalen\,der\,Physik (Leipizig) {\bf 437}, 87-102 (1948).

\bibitem{GHS2013}
K.\,Y. Bliokh and A. Aiello,
{\em Goos-H\"anchen and Imbert-Fedorov beam shifts: an overview},
J.\,of\,Opt. {\bf 15}, 014001-16 (2013).

\bibitem{2013JMO}
M.\,P. Ara\'ujo, S.\,A. Carvalho and S. De Leo,
{\em The frequency crossover for the Goos-H\"anchen shift},
J.\,of\,Mod.\,Opt. {\bf 60}, 1772-1780 (2013).

\bibitem{2014JO16}
M.\,P. Ara\'ujo, S.\,A. Carvalho and S. De Leo,
{\em The asymmetric Goos-H\"anchen effect},
J.\,of\,Opt. {\bf 16}, 015702-7 (2014).


\bibitem{2014PRA90}
M.\,P. Ara\'ujo, S.\,A. Carvalho and S. De Leo,
{\em Maximal breaking of symmetry at critical angles and a closed-form expression for angular deviations of the Snell law},
Phys.\,Rev.\, A \textbf{90}, 033844-11 (2014).

\bibitem{2016PRA93}
M. Ara\'jo, S. De Leo, and G. Maia,
{\em Closed form expression for the Goos-Haenchen lateral displacement},
Phys.\,Rev.\, A {bf 93}, 023801-9 (2016).


\bibitem{Wolf1999}
M. Born and E. Wolf,
{\sl Principles of optics} (Cambridge UP, Cambridge, 1999).

\bibitem{Saleh2007}
B.\,E.\,A. Saleh and M.\,C. Teich,
{\sl Fundamentals of Photonics} (Wiley \& Sons, New Jersey, 2007).

\bibitem{2016JMOTSB}
M.\,P. Araujo, S. De Leo, and M. Lima,
{\em Transversal symmetry breaking and axial spreading modification for gaussian optical beams},
J.\,Mod.\, Opt. {\bf 63}, 417-427 (2016).

\bibitem{OPO1}
L.\,A. Lugiato and Ph. Grangier,
{\em Improving quantum-noise reduction with spatially multimode squeezed light},
J.\,Opt.\,Soc.\,Am. B {\bf 14}, 225-231 (1997).

\bibitem{OPO2}
 C. Sauvan, P. Lalanne, and J.\,P. Hugonin,
{\em Slow-wave effect and mode-profile matching in photonic crystal microcavities},
Phys. Rev. B {\bf 71}, 165118 (2005).

\bibitem{OPO3}
A.\,R. Cowan and J.\,F. Young,
{\em Mode matching for second-harmonic generation in photonic crystal waveguides},
Phys. Rev. B {\bf 65}, 085106 (2002).

\bibitem{1999ODT}
R. Grimm, M. Weidem\"uller and Y.\,B. Ovchinnikov,
{\em Optical dipole traps for neutral atoms},
Advances in Atomic, Molecular and Optical Physics {\bf 42}, 95-170 (2000).

\bibitem{HGPRA2005}
T.\,P. Meyrath, F. Schreck, J.\,L. Hanssen, C.\,S. Chuu, and M.\,G. Raizen,
{\em Bose Einstein Condensate in a Box}
Phys.\,Rev. A {\bf 71}, 041604(R) (2005).

\bibitem{HG2005B}
T.\,P. Meyrath, F. Schreck, J.\,L. Hanssen, C.\,S. Chuu, and M.\,G. Raizen,
{\em A high frequency optical trap for atoms using Hermite-Gaussian beams}
Opt. Express {\bf 13}, 2843-2851 (2005).

\bibitem{KEM2006}
J. Magnes, D. Odera, J. Hartke, M. Fountain, L. Florence, and V. Davis,
{\em Quantitative and Qualitative Study of Gaussian Beam Visualization Techniques},
(http://arxiv.org/abs/physics/0605102).

\end{thebibliography}
\end{document}